\documentclass[fleqn,10pt]{wlscirep}
\usepackage[utf8]{inputenc}
\usepackage[T1]{fontenc}
\usepackage{mathrsfs}

\title{Magnetic ordering and dynamics in monolayers and bilayers of chromium trihalides: atomistic simulations approach}
\author[1]{S. Stagraczy\'nski}
\author[2]{P. Bal\'a\v{z}}
\author[1]{M. Jafari}
\author[1]{J. Barna\'s}
\author[1*]{A. Dyrda\l}

\affil[1]{Faculty of Physics, Adam Mickiewicz University in Pozna\'n, ul. Uniwersytetu Pozna\'nskiego 2, 61-614 Pozna\'n, Poland}
\affil[2]{FZU -- Institute of Physics of the Czech Academy of Sciences, Na Slovance 1999/2, 182 21 Prague 8, Czech Republic}

\affil[*]{adyrdal@amu.edu.pl}

\begin{abstract}
We analyze magnetic properties of monolayers and bilayers of chromium trihalides, CrI$_3$, in two different stacking configurations: AA and rhombohedral ones. Our main focus is on the corresponding Curie temperatures, hysteresis curves, equilibrium spin structures, and spin wave excitations. To obtain all these magnetic characteristic, we employ the atomistic spin dynamics and Monte Carlo simulation techniques. The model Hamiltonian includes  isotropic exchange coupling, magnetic anisotropy, and Dzyaloshinskii-Moriya interaction. Though the latter is relatively weak in CrI$_3$, we consider a more general case assuming also an enhancement of  Dzyaloshinskii-Moriya interaction in  the corresponding Janus structures  and by  external electric fields. An important issue of the analysis is the  correlation between hysteresis curves and spin configurations in the system, as well as formation of the skyrmion textures. 
\end{abstract}

\begin{document}

\flushbottom
\maketitle
 
\thispagestyle{empty}

\section*{Introduction}
Van der Waals magnetic (vdW) materials are currently of great interest due to expected applications in two-dimensional (2D) nanoelectronics and spintronics. These materials are built from weakly coupled 2D monolayers, and therefore they  can be relatively easily obtained in the multilayered form~\cite{WangJPCC}, down to a single monolayer. The discovery of ferromagnetism in monolayers of CrI$_3$~\cite{huang_layer-dependent_2017} and   Cr$_2$Ge$_2$Te$_6$~\cite{gong2017discovery} initiated an enormous interest in magnetic 2D vdW crystals~\cite{wang2022magnetic,materials,sarkar}. There are several groups of well-known magnetic vdW materials, including chromium trihalides CrX$_3$ (X= I, Cl, Br, I)~\cite{huang_layer-dependent_2017,kartsev_2020}, metal tribromides  MBr$_3$ (M=Mn, Cu, Fe, V), chromium-based ternary tellurides Cr$_2$XTe$_6$ (X = Ge, P)~\cite{gong2017discovery}, transition metal dichalcogenides MnX$_2$ and VX$_2$  (X = S, Se, Te) in two different phases denoted as 2H and 1T,~\cite{chhowalla2013chemistry,feng2011metallic,zhang2013dimension}, and also others.  
The number of known 2D magnetic van der Waals materials is growing rapidly, and they have various structural, electronic, magnetic, and transport  properties. There is currently extensive search for new materials, with new properties, and new perspectives for applications, as these materials  seem to be ideal for building ultra-thin (of atomic scale) nanoelectronic devices (like spin valves or nonvolatile memory elements) for, e.g., quantum computing, spintronics, magnonics, opto-electronics, and others\cite{wang2022magnetic,zhang_recent_2021}. 
However, magnetic properties of most of the  known 2D van der Waals systems still require further investigations. The magnetic ground state of 2D vdW crystals depends, among others,  on their crystallographic phase, stacking geometry, substrate, and on possible twisting of adjacent monolayers. In many cases,  the magnetic ground state survives only at low temperatures, while at higher temperatures the systems become paramagnetic. Therefore, much efforts are focused on searching for materials with critical temperatures above the room temperature.    

Another group of 2D materials are layered molecular magnets. Indeed, a number of molecular magnets have been synthesized in the monolayer and bilayer forms~\cite{PhysRevB.78.174409,acs.inorgchem.1c00432,acs.inorgchem.7b01930}. Both 2D  vdW materials and 2D molecular magnets offer a unique opportunity to  test various theoretical models of magnetic interactions and magnetic physical phenomena in two dimensions. As for the interactions, this includes exchange symmetric coupling, Dzyaloshinskii-Moriya Interaction (DMI), magnetic anisotropy, higher-order interactions, like two-ion magnetic anisotropy (or Kitaeev term), and also others~\cite{kartsev_2020,jaeschke-ubiergo_theory_2021}. Many of these materials have the ground state spin orientation perpendicular to the monolayers or bilayers. In some cases, the magnetic anisotropy enforces in-plane spin orientation. Moreover, these materials also allow to observe various topological phases and phase transitions, like strictly 2D skyrmions or topological Berezinsky-Kosterlitz-Thouless phase transitions~\cite{olsen_theory_2019,PhysRevB.78.174409}.       

Among various groups of vdW materials, there are materials which are  promising for applications, like for instance transition metal dichalcogenides, especially Vanadium  dichalcogenides  VX$_2$ (X=S, Se, Te)~\cite{chhowalla2013chemistry,feng2011metallic,zhang2013dimension}, and chromium  trihalides, e.g., chromium iodine and chromium chlorine, CrI$_3$ and CrCl$_3$
~\cite{huang_layer-dependent_2017,kartsev_2020,jaeschke-ubiergo_theory_2021}. In this paper we will focus on the latter group, specifically on CrI$_3$. Monolayers of this material are ferromagnetic, with the magnetic moments oriented perpendicularly to the layer plane~\cite{huang_layer-dependent_2017,li_single-layer_2020}. This material is very interesting for  fundamental research~\cite{Lado_2017,Dupont_PRL}. However, the corresponding Curie temperature is around 40 K~\cite{huang_layer-dependent_2017}, which  is too low for most of practical applications. In the bilayer form, the two monolayers are coupled antiferromagnetically. However  magnetic properties can be externally tuned by electric field~\cite{Morell_2019,ghosh_structural_2019} or strains~\cite{Ebrahimian,Leon_2020,xu_effect_2021,zheng_tunable_2017,ghosh_exotic_2022,wu_strain-tunable_2019,li_pressure-controlled_2019}, and interlayer coupling can be even changed from antiferromagnetic to ferromagnetic one~\cite{zheng_tunable_2017}. Moreover, magnetic properties of CrI$_3$ in a multilayer form depend on the stacking geometry~\cite{guo_layer_2020,acs.nanolett.9b04282,jiang_stacking_2019}. This material seems to be ideal  for testing various theoretical models of spin configuration, including topological skyrmion textures~\cite{Behera_skyrmmion,xu_magnetic_2020,Liu_AIP_ADV,Gosh_APL}, as well as excited states (spin waves) which display certain topological features~\cite{kartsev_2020,jin_raman_2018,aguilera_topological_2020,zhang_gate-tunable_2020,hidalgo-sacoto_magnon_2020,cenker}. It was pointed out~\cite{Gosh_APL}, that DMI in CrI$_3$ is too small to generate skyrmion textures. However, it was also predicted that DMI in Janus monolayers of chromium trihalides Cr(I,X)$_3$ can be significantly larger than that in CrI$_3$, and can generate magnetic skyrmion textures~\cite{xu_magnetic_2020}. Therefore, we assume a more general case, which admits larger values of DMI (e.g. in Janus structures).

\begin{figure}[ht]
\centering
\includegraphics[width=1 \textwidth]{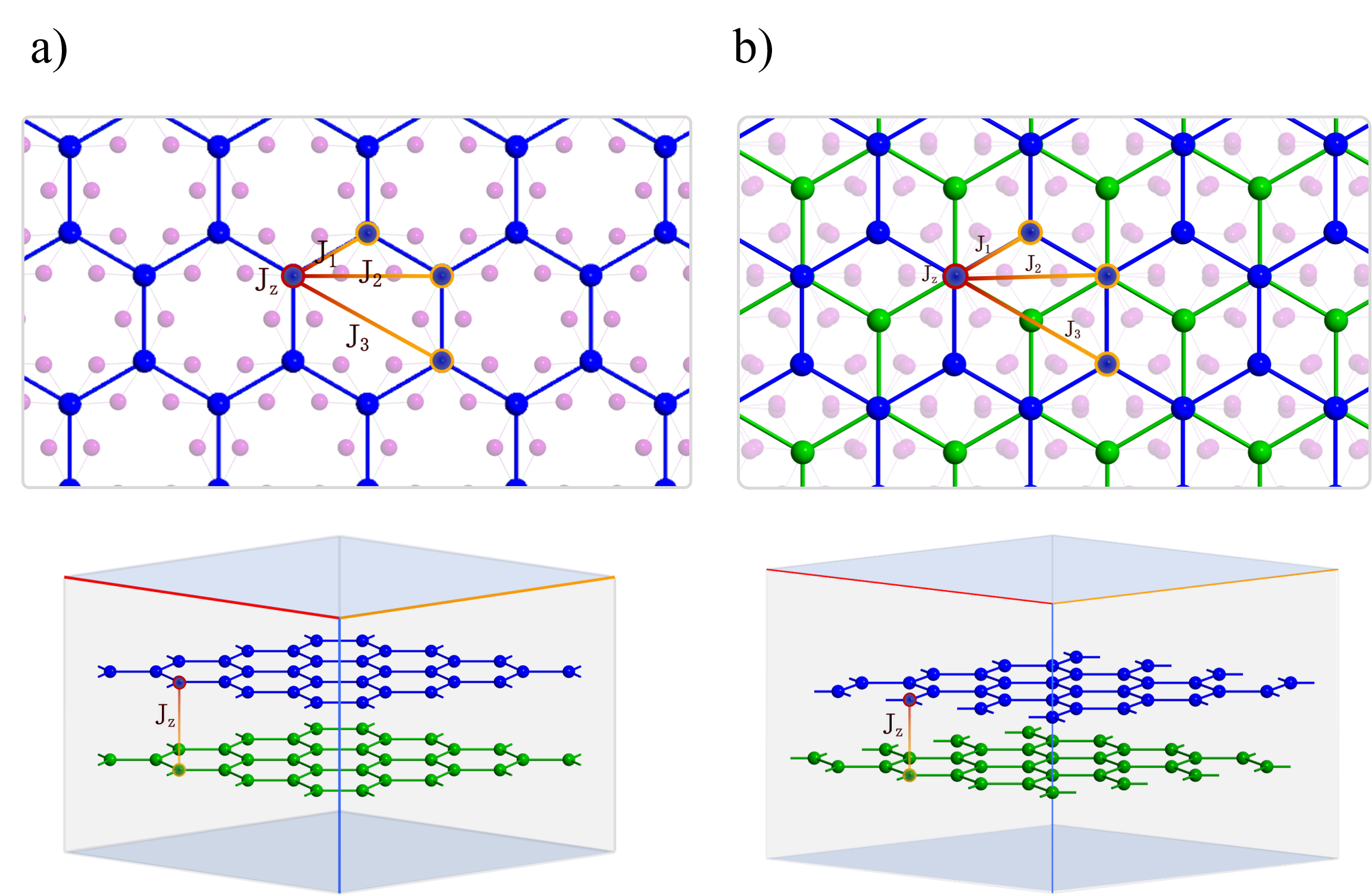}
\caption{Atomic structure of the bilayers of CrI$_3$ in the AA stacking of layers without offset (a), and  
in the  rhombohedral stacking, where the offset is half of the unit cell vectors $\vec{a} / 2$, $\vec{b} / 2$ (b). The $z$-component of unit vectors is kept the same for both cases. The Cr atoms of the top (bottom) monolayer are indicated in blue (green). The top panels present the top view, while the bottom panels show the side view. Note, the top part of (a) shows effectively atomic structure of a monolayer. The I atoms are indicated in the top parts in pink, while for clarity reasons they are not indicated in the bottom parts. Intralayer and interlayer exchange couplings are also marked in the top (bottom) parts.  }
\label{fig:geometries}
\end{figure}

In this paper we analyze magnetic properties of CrI$_3$, and to do this we employ simulation methods like atomistic spin dynamic (ASD) technique and Monte Carlo simulations~\cite{Evans_2014,Vampire_MonteCarlo,Vampire_Temperature}. From this we determine such magnetic properties like Curie temperature, hysteresis curves, (quasi)equilibrium spin configurations, or excited states (spin waves). We limit considerations to monolayers and bilayers of CrI$_3$. As some results on the monolayers are already available  in the literature, much less work has been done on the bilayer structures of  CrI$_3$. 
Atomic structure of the bilayers of CrI$_3$ is shown in Fig.{\ref{fig:geometries}} for two different stacking of the monolayers. This figure shows top and side views. Atomic structure of a single  monolayer is effectively shown by the top part of Fig.{\ref{fig:geometries}}(a).     

\section{Model and Method}

\subsection{Structure and spin Hamiltonian}

Magnetic properties of chromium trihalides remarkably depend  on the number of monolayers in the sample, and also on the geometry and offset of their stacking. This difference is especially significant for small numbers of layers in the system.  The most stable stacking geometries are the AA and Rhombohedral ones, see Fig.{\ref{fig:geometries}}. The following analysis is limited to both monolayers and bilayers of CrI$_3$. Magnetic moments in CrI$_3$ are predominantly localized on the chromium atoms, while contributions from iodine atoms are rather small and may be omitted. However, the iodine atoms contribute to exchange interactions between the magnetic Cr atoms (e.g. via superexchange), and thus have a significant impact on magnetic properties of  CrI$_3$. The Cr atoms in an individual monolayer form a hexagonal (honeycomb) lattice, and positions of these atoms have been taken from the unit cell of bulk CrI$_3$.

Accordingly, magnetic properties of CrI$_3$ will be described by an effective spin Hamiltonian attributed to the corresponding hexagonal lattice of Cr atoms, which we split into two terms, 
\begin{equation}
    \mathcal{H}= \mathcal{H}_0 + \mathcal{H}_{\rm DM}, 
    \label{eq:Hamiltonian}
\end{equation}
where $\mathcal{H}_0$ includes the exchange interactions, magnetic anisotropies, and the Zeemann energy in an external magnetic field $\mathbf B$.  The second term, $\mathcal{H}_{\rm DM}$, takes into account antisymmetric exchange, known also as Dzyaloshinskii-Moriya Interaction (DMI). 

Assuming isotropic  exchange interactions, we  write the first term in the form 
\begin{eqnarray}
  \mathcal{H}_0 = -\sum\limits_{ij} J_{ij} {\mathbf S}_i \cdot {\mathbf S}_j - k\sum\limits_i \left(S^z_i\right)^2 - \mu_S \sum\limits_{i} {\mathbf{B}}\cdot {\mathbf{S}}_i,
  \label{eq:SpinHamiltonian}
\end{eqnarray}
where the sum over $i$ and $j$ means the summation over lattice sites.  
Here, the positive  exchange parameters $J_{ij}$ correspond to ferromagnetic (FM) coupling, whereas negative ones correspond to antiferromagnetic (AFM) coupling. In the following we take nonzero exchange parameters between nearest-neighbours, second-nearest and third-nearest neighbors. In a more general case the exchange parameters can be anisotropic. The second term describes the magnetic anisotropy. Here, positive  constant $k$ corresponds to the  easy-axis anisotropy along the $z$-direction (normal to the layer), while negative $k$ corresponds to the easy-($xy$)plane anisotropy. Furthermore, $\mu_s$ is the atomic magnetic moment of Cr atoms. Here, and in the following, if not stated otherwise, the vector $\mathbf{S}$ is a unit vector along the orientation of a local spin moment~\cite{Evans_2014}. Accordingly, one needs to adjust  the exchange parameters, which usually are related to spin Hamiltonian, where $\mathbf{S}$ is the spin operator.     

The second term in Eq.(\ref{eq:Hamiltonian}) includes the  Dzyaloshinskii-Moriya interactions (or alternatively antisymmetric exchange). Existence of nonzero components of this interaction depends on symmetry of the system. In van der Waals materials one can easily tune externally the symmetry and thus also the components of DMI, e.g., by an external mechanical strain or by electric field normal to the layers. In the case under consideration, the DMI  Hamiltonian can be written in the form~\cite{jaeschke-ubiergo_theory_2021}  
\begin{equation}
    \mathcal{H}_{\rm DM} =    -\sum\limits_{\left\langle i, j \right\rangle} {\mathbf{d}}^{NN}_{ij} \cdot \left( \mathbf{S}_i \times \mathbf{S}_j \right)          -\sum\limits_{\left\langle\left\langle i, j \right\rangle\right\rangle} \mathbf{d}^{NNN}_{ij} \cdot \left( \mathbf{S}_i \times \mathbf{S}_j \right),
    \label{eq:DM_Hamiltonian}
  \end{equation}
where the first term includes interaction between intralayer nearest-neighbours (NN), while the second term between next nearest-neighbours (NNN). Here, ${\mathbf{d}}^{NN}_{ij}$ and ${\mathbf{d}}^{NNN}_{ij}$ are the corresponding Dzyaloshinskii-Moriya vectors~\cite{jaeschke-ubiergo_theory_2021}.      
This vector for NN has the following form:  $\mathbf{d}^{NN}_{ij} = d^{NN}_{xy} \hat{\mathrm{z}} \times  \hat{\mathrm{l}}_{ij} + \tau_{ij}^{NN} d^{NN}_z \hat{\mathrm{z}}$, where $\hat{\mathrm{z}}$ is a unit vector along the axis $z$ (normal to the layer), $\hat{\mathrm{l}}_{ij}$ is the unit vector oriented from the site $i$ to site $j$, while $d^{NN}_{xy}$ and $d^{NN}_z$ are parameters  and $\tau_{ij}^{NN} =1 (-1)$  for the AB (BA) link. 
In turn, in the second term 
$\mathbf{d}^{NNN}_{ij} = \tau^{NNN}_{ij} \left( d_{xy}^{NNN} \hat{\mathrm{l}}_{ij} + \nu_{ij} d_z^{NNN} \hat{\mathrm{z}} \right)$, with $\tau^{NNN}_{ij} = +1$ for AA-link  and $-1$ for BB-link, while   $\nu^{NNN}_{ij} = \pm 1$ and alternates its sign on the consecutive NNN links of a chosen site~\cite{jaeschke-ubiergo_theory_2021}. 

The DMI between nearest-neighbors in pristine CrI$_3$ vanishes, while the DMI  between the next-nearest-neighbors is  rather small, $d_z^{NNN}=-8.8\mu\mathrm{eV}$ and $d_{xy}^{NNN}=73.0\mu\mathrm{eV}$~\cite{jaeschke-ubiergo_theory_2021}.
However, DMI can be tuned by an external electric field normal to the layer, so the parameters $d_{xy}$ and $d_z$ can be nonzero and relatively large~\cite{jaeschke-ubiergo_theory_2021}. It can be also tuned by external strain. Moreover, the DMI can be  significantly enhanced in the corresponding  Janus monolayers of chromium trihalides Cr(I,X)$_3$~\cite{xu_magnetic_2020}.  All this makes the DMI parameters between NNs and NNNs tunable in a quite considerable range. Finally,we 
note, that in the Hamiltonian assumed here we neglect such  terms, like anisotropic exchange and higher-order anisotropy terms (e.g., two-ion anisotropy or Kitaev term).

\subsection{Numerical procedures}


Our numerical calculations are based on the  atomistic spin dynamics (ASD) method implemented into the Vampire code~\cite{Evans_2014}. This method allows for taking into account all details of the system geometry, and also for keeping insight into behaviour of the spin moments during the simulation process. The ASD simulations are carried out by solving the coupled system of atomistic stochastic (at finite temperature) Landau–Lifshitz-Gilbert (LLG) equations, 
\begin{equation}
\label{eq:LLG}
\frac{\partial \mathbf{S}_i}{\partial t} = -\frac{\gamma}{1 + \alpha^2} \left[\mathbf{S}_i \times \mathbf{B}^\mathrm{eff}_i + \alpha \mathbf{S}_i\times (\mathbf{S}_i \times  \mathbf{B}^\mathrm{eff}_i )\right],
\end{equation}
where $\gamma$ is the gyromagnetic ratio, $\alpha$ stands for the Gilbert damping parameter, and $\mathbf{B}^\mathrm{eff}_i$ is the effective magnetic field acting on the spin at site $i$.
This effective field $\mathbf{B}^\mathrm{eff}_i$ consists of two terms: one is deterministic and is related to the spin Hamiltonian, while the second one is a stochastic term, that is determined by thermal fluctuations, 
\begin{equation}
    \mathbf{B}^\mathrm{eff}_i = \frac{-1}{M_s} \frac{\partial \mathscr{H}}{\partial \mathbf{S}_i} + \mathbf{\xi}^{\rm (th)}_i. \label{EffectiveField}
\end{equation}
In the low energy regime, the stochastic term, $\mathbf{\xi}^{\rm (th)}_i$ is usually modelled by the Gaussian thermal noise.

To verify the results on spin dynamics obtained by the atomistic LLG equations, we also  used other methods,  like the method based on the spin-spin correlation function and  the  Monte Carlo simulation technique. We find a very good agreement between the results obtained by the atomistic LLG equations and those obtained by the Monte Carlo simulations. As the former technique is more appropriate for dynamical properties, the latter one is more convenient for static properties. 
However, the static properties are also accessible from spin dynamics, but require much higher computational efforts.  In addition, to study dynamical properties we also use the atomistic approach implemented within the Uppsala ASD code~\cite{Skubic_2008}.
 
 An important step in the simulations process, especially when simulating    
 hysteresis loops, critical temperatures, or magnon dispersion relations, is the preparation of  the relevant initial spin configuration. To do this we use the field-cooling procedure. 
 We  start this procedure always with the random spin state at a high temperature (above the Curie temperature). After performing one million of Monte Carlo equilibration steps (or time steps for LLG equation), the temperature becomes decreased nearly adiabatically towards the minimum temperature (i.e. temperature, at which the required simulations are to be performed, e.g. 15 K in most of our simulations of the hysteresis loops). 

 In the following we will apply the above mentioned methods to determine the Curie (N\'{e}el) temperature, as well as hysteresis loops at a finite (nonzero) temperature. The simulations will be performed for monolayers, as well as for bilayers in two different stacking geometries, i.e., in the AA and rhmbohedral configurations. In addition we will also consider dynamical properties of the systems under consideration, especially spin wave excitations.    
 
\section{Results} 

\subsection{Critical temperatures}
The reported experimental data indicate on a ferromagnetic  ground state in CrI$_3$ monolayer~\cite{huang_layer-dependent_2017}. Moreover, FM ordering  is also expected in multilayers of chromium trihalides, (i.e. in corresponding bulk material). However, the bilayer of CrI$_3$  was reported to display antiferromagnetic interlayer exchange coupling~\cite{huang_layer-dependent_2017}, so such bilayers should also reveal not only features typical of ferromagnets, but also some properties typical of antiferromagnets.
The interlayer coupling can be tuned (including also sign change), e.g., by an external gate voltage~\cite{Morell_2019,ghosh_structural_2019} or strain\cite{Leon_2020,xu_effect_2021,zheng_tunable_2017,ghosh_exotic_2022,wu_strain-tunable_2019,li_pressure-controlled_2019}.  Therefore, one of our main objectives in the case of bilayers is to  investigate the impact of the layer stacking on the critical temperature. The analyzed geometries (AA and rhombohedral) are visualized in Fig.~\ref{fig:geometries}. 
Second, in antiferromagnetically coupled bilayers one may expect certain features related to the breakdown of ferromagnetic order within the monolayers and of antiferromagnetic order related to the interlayer coupling (Curie and N\'{e}el temperatures.
    
\begin{figure}[htb]
    \centering
    \begin{picture}(\textwidth,2.35in)
        \put(0,0){\includegraphics[width=0.45\textwidth]{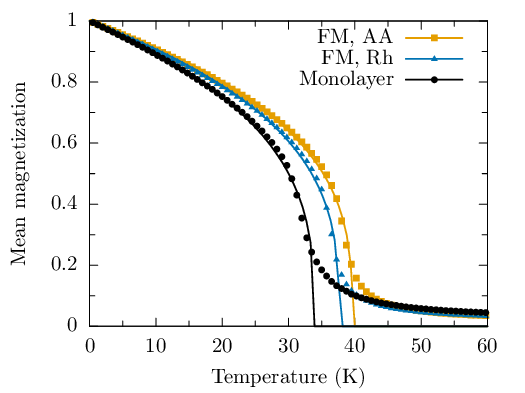}}
        \put(0.5\textwidth,0){\includegraphics[width=0.45\textwidth]{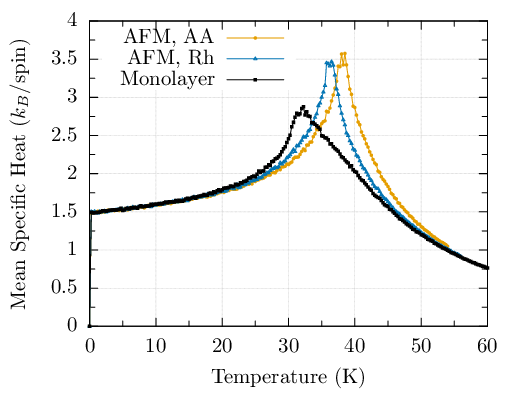}}
        \put(0.52in,2.32in){\bf a)}
        \put(3.98in,2.32in){\bf b)}
    \end{picture}
\caption{(a) Simulated (dots) magnetization vs. temperature for the monolayer and ferromagnetic bilayers in the AA and Rh-stacking. The Curie temperatures have been determined from fitting to the  formula $(1-T/T_C)^\beta$ (solid lines). From this fitting follows: {monolayer: $T_C = 33.93$K, $\beta=0.32$; FM-coupled bilayer in  Rhombohedral stacking: $T_C = 37.64$K, $\beta=0.31$; FM-coupled bilayer in the AA stacking: $T_C = 39.64$K, $\beta=0.32$. 
   (b) Simulated mean specific heat vs. temperature for the  antiferromagnetic bilayers in the AA and Rh-stacking.. We added here also the data for the ferromagnetic monolayer, to show that the Curie temperatures determined from simulations of the  specific heat and from simulations of the magnetization are equal. 
       The parameters assumed in simulations:  The exchange couplings (after Ref.~\citeonline{kartsev_2020})  
    $J_1 = 4.020$meV, $J_2 = 0.643$meV, $J_3 = 0.016$meV), while for the interlayer exchange 
    in bilayers we assumed $J_z = J_2$ (a) and $J_z = -J_2$ (b).  
    The nearest-neigbour DMI, 
    $d^{NN}_{xy} = 0.201$meV and $d^{NN}_z=0$,  the magnetic anisotropy constant $k = 0.109$ meV, and the chromium magnetic  moment $\mu_{Cr} = 3.36 \mu_B$. The parameters of the NNN DMI as in the text.} }  
    \label{fig:Tc}
    \end{figure}
    
The simulated data (points) for both monolayer and ferromagnetic bilayer (ferromagnetic interlayer coupling) in the AA and Rh stacking geometries are shown in Fig.~\ref{fig:Tc}(a). We included there  the exchange parameters between nearest neighbours, $J_1$, second-nearest neighbours, $J_2$, third-nearest neighbours, $J_3$, and interlayer exchange coupling between nearest neighbours, $J_z$, as described in the figure caption. The other parameters are also described there.       
To determine the  Curie temperatures, the corresponding data have been fitted to the formula $(1-T/T_C)^\beta$, and from this fitting we found $T_C \approx 33.93$ K for the monolayer, $T_C \approx 37.64$ K for the Rh-stacked bilayer, and $T_C \approx 39.64$ K for the AA-stacked bilayer. The Curie temperature for the monolayer is lower than that for the bilayers, and the bilayer in the AA-stacking has Curie temperature higher than that for the Rh-stacking. For the assumed parameters, the DMI has no significant impact on the magnetization vs. temperature profile and on Curie temperature. 

The critical temperatures (both Curie and N\'{e}el) also can be determined from magnetic susceptibility or specific heat. In the case of ferromagnetic monolayer and ferromagnetically coupled bilayers, both these methods give results which are consistent with those obtained from the above magnetization analysis. Therefore, to determine the critical temperature in bilayers with antiferromagnetic interlayer coupling, we  used the  Vampire package~\cite{Vampire_Temperature,Vampire_MonteCarlo} to simulate the specific heat as a function of temperature, $ C_v = [\left\langle U^2 \right\rangle - \left\langle U \right\rangle^2]/k_B T^2$, where $U$ is the internal energy. The specific heat has a cusp at the critical temperature, and this property is used to determine the Curie temperature, $T_c$, in ferromagnetic systems, and the N\'{e}el  temperature in antiferromagnets. 

The numerical results on the specific heat for  bilayers with antiferromagnetic coupling in the AA and Rh stackings and also for the ferromagnetic monolayer 
are shown in Fig.~\ref{fig:Tc}(b). The critical temperatures  for the bilayers with antiferromagnetic interlayer coupling are clearly seen and are roughly the same as Curie temperatures for the corresponding bilayers with ferromagnetic interlayer coupling. This is because absolute value of interlayer coupling is remarkably smaller than the intralayer exchange coupling constant $J_1$, and absolute magnitude of the interlayer coupling is assumed the same in bilayers with ferro- and antiferromagnetic interlayer coupling. The curve for ferromagnetic monolayer is presented in Fig.~\ref{fig:Tc}(b) just to show that the methods used in simulations of the results in Fig.~\ref{fig:Tc}(a) and Fig.~\ref{fig:Tc}(b)  give the same critical temperatures. 

When the interlayer exchange coupling is antiferromagnetic, the total magnetic moment of the bilayer vanishes in the ground state. As already mentioned above, the absolute magnitude of the interlayer coupling is remarkably  smaller than the intralayer exchange coupling $J_1$. Thus, with increasing temperature, the antiferromagnetic moment (N\'{e}el vector) vanishes first and we have a system of two ferromagnetic layers (with average moments either aligned or antialigned). When the temperature grows further the magnetic moment in each monolayer (and thus in the whole system) drops to zero at the Curie temperature as determined above. Thus, on the simulated magnetization vs temperature curves one could expect two cusps, one at a lower temperature that corresponds to the breakdown of the antiferromagnetic alignment of the two monolayers, and another one at higher Curie temperature. However, in the simulation data on the magnetization vs temperature, there is no signature of the breakdown of antiferromagnetic configuration. Apparently, it would be resolved in a multilayer system~\cite{curieneel}. 

\begin{figure}[htb]
        \centering
    \includegraphics[width=0.6\textwidth]{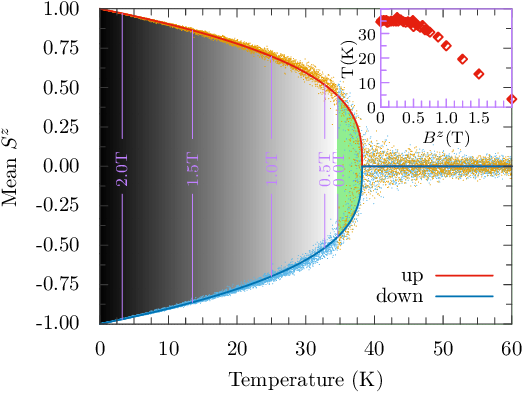}
        \caption{Simulated normalized values of $S^z$ for both monolayers in the AFM-coupled  Rh-stacked bilayer, presented as a function of temperature. These normalized values of $S^z$ for the top and bottom layers are shown as the yellow and blue  dots, respectively. The solid red and blue lines indicate the average of $S^z$ for both layers.  The grey  area between these lines indicates the temperature range where reorientation of the magnetic ordering between the layers does not  happen in the absence of external magnetic field. Reorientation begins at the border between the grey and green areas, and the temperature at which this takes place (around 34 K) we assign as the N\'{e}el temperature $T_N$. When an external magnetic field normal to the bilayer is applied, the reorientations start at lower temperature, as shown in the inset, and also marked schematically on the grey area. }
          \label{fig:Sz_T_norm}
    \end{figure}

In order to get some information on the temperature at which the antiferromagnetic alignment of the two ferromagnetic monolayers is destabilized by thermal fluctuations, we analyzed average values of the total spin $z$-component of each monolayer separately as a function of temperature.
Results of the simulations are shown in Fig.~\ref{fig:Sz_T_norm}, where the simulated normalised values of $S^z$ for the two monolayers are presented  with blue and yellow dots, while the average values are shown by the solid blue and red lines.  With increasing temperature, fluctuations around the average values become larger, and at a certain temperature (recognized as the relevant N\'{e}el temperature) the average spin moments of the two monolayers start to change sign. The N\'{e}el temperature was identified as a temperature at which spin moments of the monolayers begin to fluctuate between the up and down orientations, see Fig.~\ref{fig:Sz_T_norm}. The corresponding  N\'{e}el temperature is lower than the Curie temperature by about 3.5 K, as one might expect.    

\subsection{Hysteresis loops and spin textures}

Hysteresis curves are important characteristics as they include information on magnetization variation with increasing external magnetic field and thus also on possible field-induced phase transitions. 
We have determined the hysteresis loops from the Monte-Carlo (MC) simulations as well as from the atomistic Landau-Lifshitz-Gilbert (LLG) equation. The starting point for each simulation was a random state at temperature $T=80\mathrm{K}$ and cooling field $H_{\mathrm{fc}}=1\mathrm{T}$ applied perpendicularly to the sample plane. The field-cooling procedure was used to cool the system's temperature down to $T=15\mathrm{K}$, and then the simulations of a hysteresis loop was initiated.
 
\subsubsection{Monolayers}
    
\begin{figure*}[t]
    \centering
    \includegraphics[width=0.89\textwidth]{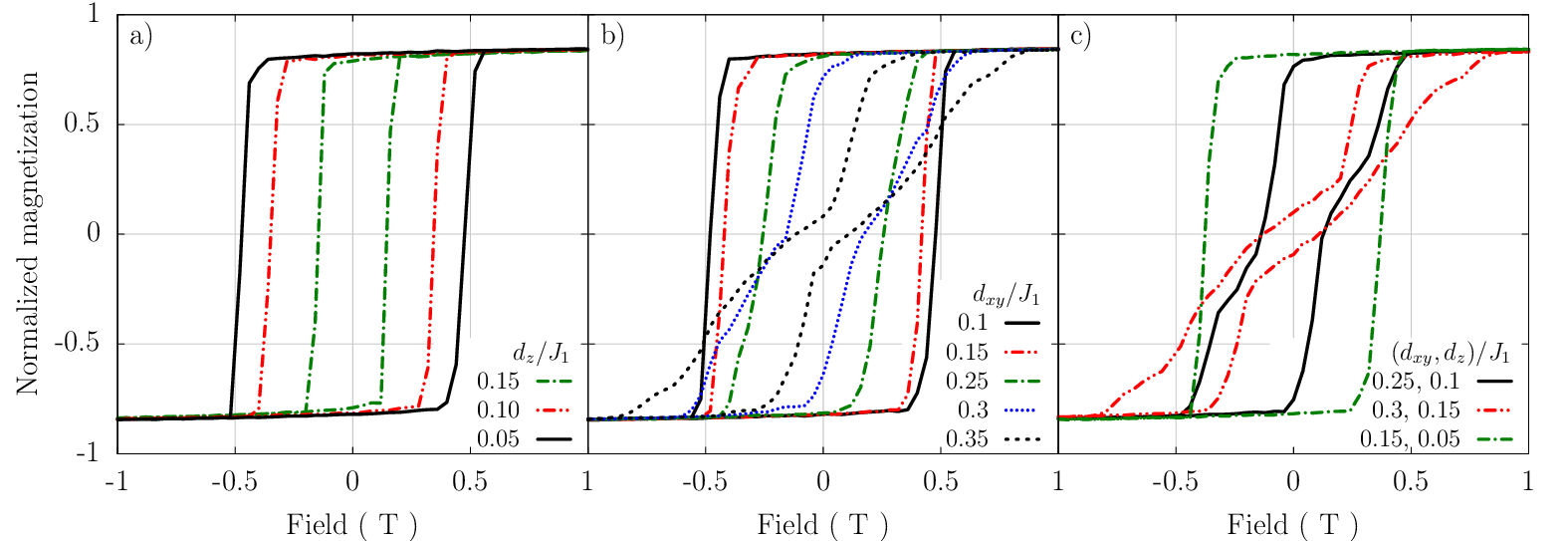}
    \caption{ Hysteresis loops for monolayers at $T=15$K,  easy-axis (normal to the layer) anisotropy  parameter $k = 0.109$meV, and with the NN DMI parameters: (a) $d_{xy}=0$ and $d_z/J_1$ as indicated; 
    (b) $d_{z}=0$ and $d_{xy}/J_1$ indicated; and (c) ($d_{xy}/J_1$, $d_z/J_1$) as indicated. 
    The other parameters  as  in Fig.~\ref{fig:Tc}. The presented  results are  average  over 50 realizations.} 
    \label{fig:monolayers}
\end{figure*}
 
Figure~\ref{fig:monolayers} shows the hysteresis curves for fixed values of DMI between NNNs (as described above in the text), and different values of the  components $d_z$ and $d_{xy}$ of the DM vector between NNs. When $d_{xy}=0$ , then increasing the component $d_z$ leads mainly to reduced  coercivity of the hysteresis curves, while their shape remains similar to those for ferromagnetic layers, see Figure~\ref{fig:monolayers}(a). 
A more interesting impact of DMI on the hysteresis curves appears for the $d_{xy}$ component of the DMI vector. The corresponding hysteresis curves for monolayers are shown in  Figure~\ref{fig:monolayers}(b)  for different values of  $d_{xy}$ and $d_z=0$. For small values of $d_{xy}/J_1$, the hysteresis curves are qualitatively similar to those for zero $d_{xy}$, but the corresponding coercivity becomes reduced with increasing $d_{xy}$. When $d_{xy}$ grows further, some new features appear in the curves, see e.g. the curves for $d_{xy}/J_1=0.3$ and $d_{xy}/J_1=0.35$, which clearly distinguish these loops from those typical for ferromagnetic systems.  
The interplay of both components of NN DMI vector is presented in Figure~\ref{fig:monolayers}(c), 
where both components are nonzero and comparable. 
For the assumed parameters the  corresponding hysteresis loops  contain clearly  features of both  $d_z$  and $d_{xy}$ components of the DMI vector.    
From the hysteresis curves shown above one may conclude that DMI modifies magnetic texture of the system and this modification leads to the observed features in the hysteresis loops. In the curves shown here, the most pronounced features are for $d_{xy} =0.35J_1$. Therefore we analyse now in more details this specific situation ($d_z=0$ and $d_{xy} =0.35J_1$), and for clarity, we also omit here the NNN DMI.

    \begin{figure}
    \centering
    \includegraphics[width=\textwidth]{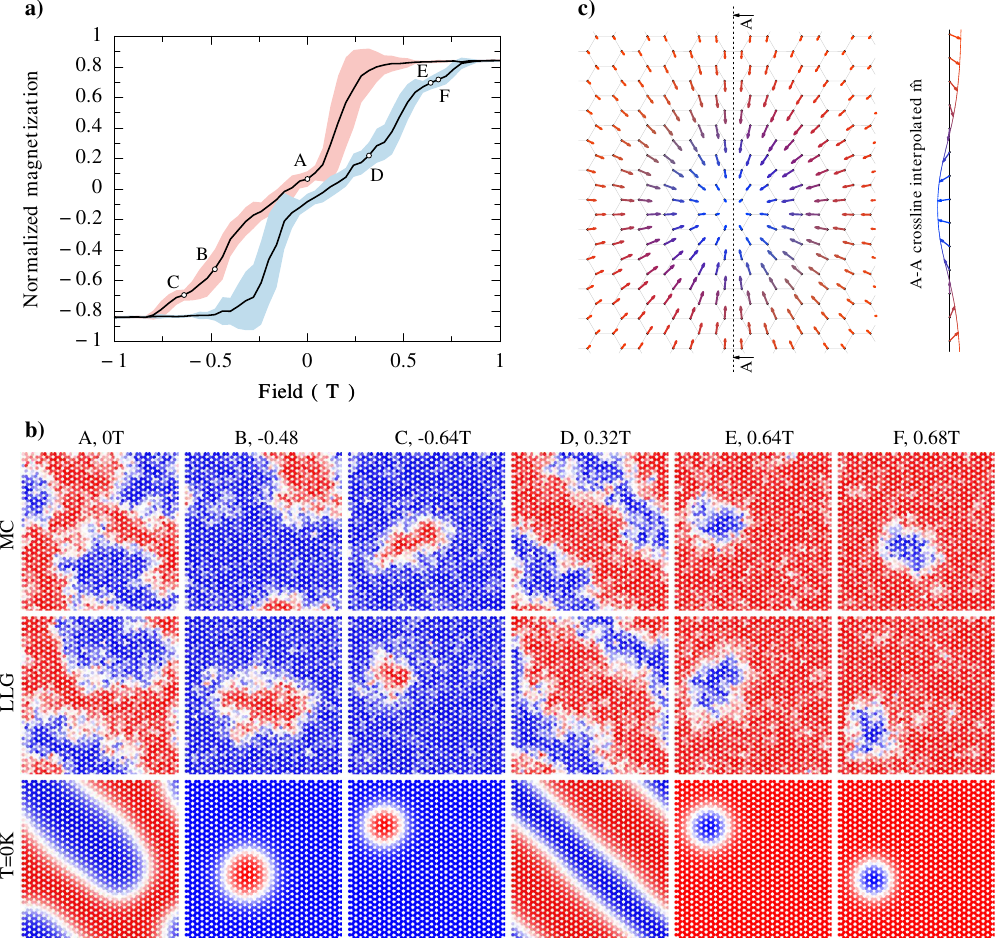}
     \caption{(a) Hysteresis loop for the monolayer with $d_{xy} = 0.35 J_1$, $d_z =0$, $T=15$K, and vanishing NNN DMI. The solid black line is the average over 50 realizations, while the shaded area represents  the standard deviation. The circle points represent realizations that are closest to the average (solid line), for which the corresponding magnetic textures  are shown in part (b) of this figure.
     (b) The magnetic textures corresponding to the points A, B, C, D, E and F, marked in the hysteresis loop (a). The results of MC simulations are shown in the top row. Then, the final set of MC simulations was used as initial step for simulations based on the atomistic LLG equation, and the middle row shows the magnetic textures after $1$ns of spin-dynamics using the LLG Heun method with damping factor $\lambda=0.1$. Then the results from the middle row were used in the next simulation for cooling the temperature from $15$K down to  $0$K. The spin textures upon cooling down to 0K are shown in the bottom row. Upon cooling, the magnetic patterns (stripes, skyrmions) have regular edges.  
     (c) Magnified single skyrmion obtained upon cooling the magnetic pattern corresponding to the point F ($B=0.68\mathrm{T}$) in (b). Here, the spin structure of the skyrmion is clearly resolved. This spin structure also confirms that the created skyrmions are of Néel-type, as presented by the vertical crossection through the skyrmion center, shown on the right side.}
        \label{fig:hyst-loops-15K}
   \end{figure}

    In Figure~\ref{fig:hyst-loops-15K}(a) we show the average hysteresis curve  (solid line) and  the standard deviation from the average (shaded areas)  
    for $d_{xy} =0.35J_1$, $d_{z} =0$ and 15 K.  
    The corresponding spin patterns for the points A, B, C, D, E and F, indicated on the hysteresis curve are shown in Figure~\ref{fig:hyst-loops-15K}(b). 
%
    Here, the results of MC simulations are shown in the top row. Then, the final set of MC simulations was used as initial step for simulations based on the atomistic LLG equation. The middle row shows the magnetic textures after $1$ns of spin-dynamics using the LLG Heun method with damping factor $\lambda=0.1$. We have changed the numerical integrator from Monte Carlo to LLG in order to check if the spin texture was stable under both integrators. 
    The results from the middle row were subsequently used in the next simulation for the cooling temperature from $15$K down to  $0$K.
    As the patterns for $T=15$K  have spin textures irregular at the boundaries due to thermal fluctuations, the edges become regular when cooling the system temperature down to 0 K, see  bottom row in Figure~\ref{fig:hyst-loops-15K}(b). As one can see, upon cooling the edges of various objects become smooth and one can clearly distinguish there regular circular objects and stripe structures. The circular objects are N\'{e}el-type  skyrmions. To see this in more details, in  Figure~\ref{fig:hyst-loops-15K}(c) we zoomed out an individual circular object, so now the spin structure associated with  the skyrmion is  clearly visible. To show further details, on the right side of this figure we present a crossection through the center of the  skyrmion . 

\subsubsection{Bilayers}

\begin{figure}
       \centering
       \includegraphics[width=\textwidth]{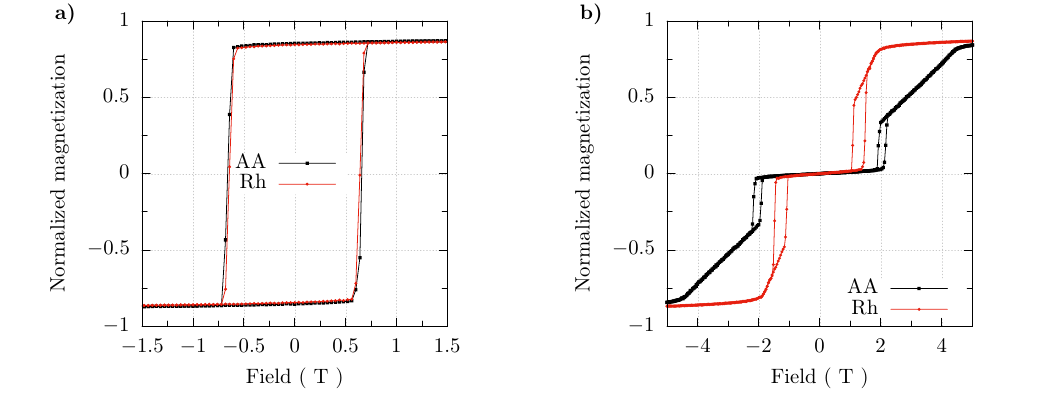}
        \caption{Hysteresis loops for bilayers in the AA and Rh stacking with (a) ferromagnetic interlayer coupling  $J_z = J_2$, and (b) antiferromagnetic interlayer coupling, $J_z =-J_2$. The out-of-plane anisotropy constant $k = 0.109$ meV, and 
                temperature $T=15$K. NN and NNN terms of DMI are omitted here. }
        \label{fig:XXX_FM}
        \label{fig:XXX_AFM}
    \end{figure}

Hysteresis loops in bilayers  depend  of the nature of interlayer coupling, and for ferromagnetic and antiferomagnetic couplings are usually remarkably different. In addition, they also depend on the stacking geometry. The hysteresis curves for the AA and Rhombohedral stackings in the absence of DM interactions and for isotropic exchange interactions are shown in Fig.\ref{fig:XXX_FM}(a) for  systems with ferromagnetic interlayer coupling and in Fig.\ref{fig:XXX_AFM}(b).
for systems with antiferromagnetic interlayer coupling. 
    \begin{figure}
    \centering
    \includegraphics[width=\textwidth]{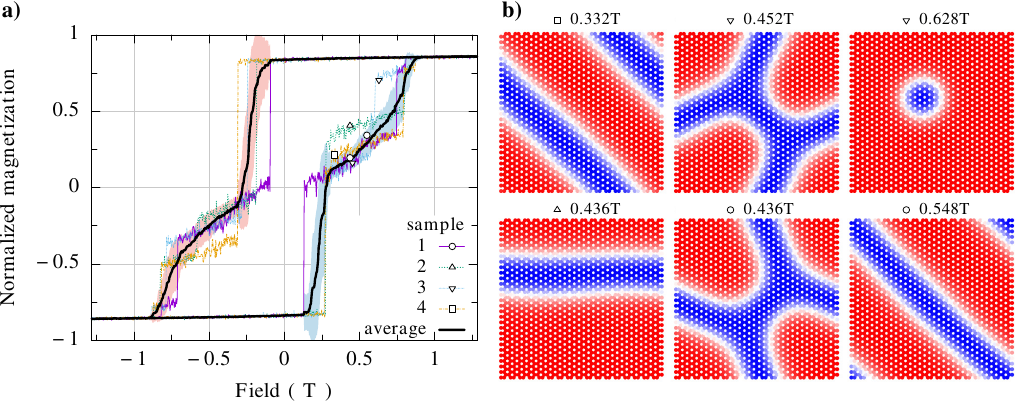}
    \caption{(a) Hysteresis loop of a ferromagnetically coupled bilayer in the Rh-stacking, calculated for $d_{xy} = 0.35 J_1$,  $T=15$K, ferromagnetic interlayer coupling $J_z=J_2$, and  for constant NNN DMI as described in the main text. The solid black line is the average over  50 realizations, while the shaded area represents the standard deviation. The numbers 1 to 4 indicate four exemplary hysteresis loops obtained  in the numerical simulations. (b) Spin patterns in FM-coupled Rh-stacked bilayers upon cooling, obtained for the points  indicated on the hysteresis loop in (a). Only spin pattern of the top layer is shown here, as that of the bottom layer is approximately  the same. One can distinguish here stripes, skyrmions, and other patterns with regular edges. 
    }
    \label{fig:FM-DMIxy-XXX-Rh-15K-NNN}
\end{figure}

    From this figure follows that the hysteresis curves for ferromagnetically coupled bilayers are similar to  typical hysteresis loops  of ferromagnetic layers,  and are roughly independent of the stacking geometry. On the other hand, the hysteresis curves for bilayers with  AFM interlayer coupling are significantly different from those for ferromagnetically coupled bilayers, and also different for different stacking geometries.   
    An important feature of the hysteresis loops for antiferromagnetic interlayer coupling is a wide plateau  at zero magnetization, which corresponds to antialigned magnetizations of the two monolayers. At a certain value of magnetic field there is a jump in magnetization associated with transition to the spin-flop phase (note, the magnetic field is normal to the layers). Then the system goes smoothly to the saturated state. The magnetic field, at which the transition to the spin-flop phase occurs with increasing magnetic field is different from the field at which the system goes back to antialigned configuration when the magnetic field  decreases. This leads to minor loops clearly visible in Fig.\ref{fig:XXX_AFM}(b) on both sides (positive and negative magnetic fields). 
   These loops follow from the magnetic anisotropy in the system, and also depend on the stacking geometry. No such features appear in ferromagnetically coupled bilayers. The saturation field also depends on the stacking configuration. This is because the saturation field depends on the effective coupling between the two monolayers, and this coupling depends, in turn, on the  strengths of interlayer exchange parameter, and on the number of interlayer nearest neighbours. Note, the hysteresis curves for the bilayers in the AA and Rh stacking have been calculated for the same interlayer exchange parameters. However, the numbers of nearest neighbors in the Rh stacking is larger (but distance between them is longer then in the AA stacking). Thus, for the assumed parameters the  saturation fields are different for  both stacking geometries and is larger for the Rh stacking.

\begin{figure}
    \centering
    \includegraphics[width=\textwidth]{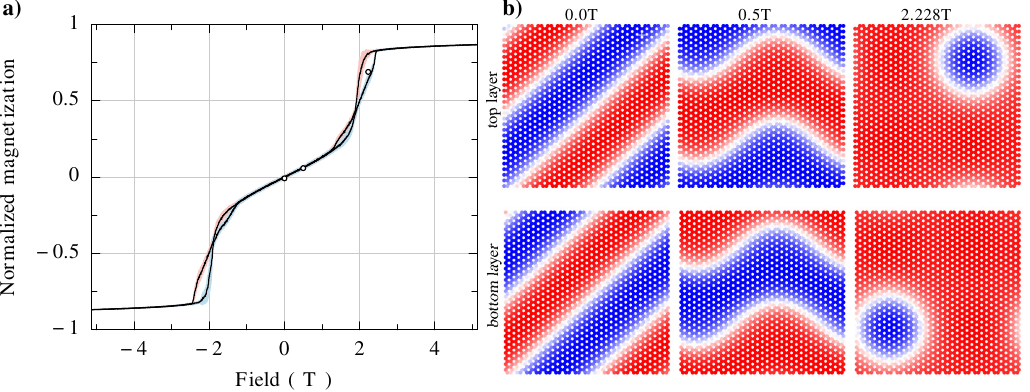}
    \caption{(a) Hysteresis loop for antiferromagnetically coupled Rh-stacked bilayer, simulated  for $d_{xy} = 0.35 J_1$, $d_z =0$, $T=15$K, antiferromagnetic interlayer coupling $J_z = - J_2$, and NNN DMI as described in the text. The solid black line is the average over 50 realizations, with the shadow represents the standard deviation. (b) The spin patterns for three values of the magnetic field, including the case with zero magnetic field. These magnetic fields correspond to the three points indicated in the hysteresis loop (a).  }
    \label{fig:hyst-loops-15K-bi-Rh-AFM}
    \end{figure}

 The impact of both NN and NNN terms of DMI  on the hysteresis curves in the Rh stacking geometry and for ferromagnetic and antiferromagnetic interlayer coupling is shown in Fig.\ref{fig:FM-DMIxy-XXX-Rh-15K-NNN}(a) and Fig.\ref{fig:hyst-loops-15K-bi-Rh-AFM}(a), respectively, while the corresponding spin textures are shown in Fig.\ref{fig:FM-DMIxy-XXX-Rh-15K-NNN}(b) and Fig.\ref{fig:hyst-loops-15K-bi-Rh-AFM}(b). As already discussed above, the DMI between NNNs is rather small in pristine CrI$_3$. In turn, the DMI between NN's can be easily tuned by external gate and therefore can be dominant (even if it vanishes in pristine CrI$_3$), so in Fig.\ref{fig:FM-DMIxy-XXX-Rh-15K-NNN}(a) and Fig.\ref{fig:hyst-loops-15K-bi-Rh-AFM}(a) we assumed intrinsic values of  NNN DMI (as given in the text) and quite significant value of DMI between NNs, $d_{xy} = 0.35 J_1$ and $d_z=0$. We focus on the impact of the component $d_{xy}$ of DMI as this component is crucial in the skyrmion texture formation.           
As in monolayers, the DMI modifies the hysteresis loops and also the corresponding spin patterns. These patterns include areas of various shapes, with irregular edges due to thermal fluctuations at finite temperatures.   However, when the temperature is cooled down to T=0K, these edges become regular and reveal among others stripe structures and well defined skyrmion textures.

Consider first the bilayers with ferromagnetic interlayer coupling. The corresponding spin patterns are shown in Fig.\ref{fig:FM-DMIxy-XXX-Rh-15K-NNN}(b)  for  a few points indicated on the corresponding hysteresis curve, see Fig.\ref{fig:FM-DMIxy-XXX-Rh-15K-NNN}(a). In general, we find there stripes, skyrmions, and also other structures. Only spin patterns of the top monolayer are shown there as the spin patterns of the bottom monolayer are practically the same due to ferromagnetic interlayer coupling. Certain, almost not resolved difference may appear due to a shift of the two monolayers in the Rh stacking. Accordingly, the  skyrmions in both layers are coupled ferromagnetically and have the same topological charge.

In bilayers with antiferromagnetic coupling, the corresponding spin patterns are qualitatively similar to those in FM-coupled bilayers, and we find the spin stripes, wavy stripes, and skyrmions, as shown in  Fig.\ref{fig:hyst-loops-15K-bi-Rh-AFM}(b). Though they look qualitatively similar to those in FM-coupled bilayers, they are different and this difference originates from different magnetic fields for which the spin patterns have been simulated and also from opposite signs of the interlayer coupling.  
IN the simulation results shown in Fig.\ref{fig:hyst-loops-15K-bi-Rh-AFM}(b), we show the spin patterns of both top and bottom layers. Note, for small magnetic fields (H=0 and H=0.5 T), the spin patterns of  both layers are practically opposite due to antiferromagnetic interlayer coupling. The situation is different for stronger magnetic field, H=2.28 T, which is strong enough to overcome te interlayer exchange coupling. In that case  the spin patterns of both layer reveal individual skyrmion in the spin pattern of the top layer and also individual skyrmion in the bottom layer. These two skyrmions, however, are not coupled and appear in different positions.  

Finally, we
also note  here, that the hysteresis loops and the corresponding spin patterns of the monolayers and bilayers, as shown above,  are for isotropic exchange interactions. These loops and spin patterns may become remarkably modified when the exchange interaction has anisotropic contributions, or more complex anisotropy terms (like two-ion anisotrop) occur in the system. A typical example  of such interactions is the so-called Kitaev term.  

\subsection{Artificially created Skyrmion lattices} 

\begin{figure}[h]
    \centering
    \includegraphics[width=\textwidth]{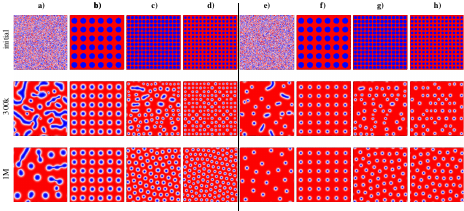}
    \caption{Spin patterns in FM-coupled Rh-stacked bilayers upon cooling for high density of skyrmions. The left part (the columns (a-d)) is for $B$=0.3T, while the right part (columns (e-h)) is for $B$=0.748T. The top row shows the initial spin configurations: random in (a) and (e) and set artificially in the others. The other two rows correspond to the spin structure of the bottom layer after certain number of Monte Carlo steps, as indicated. The other parameters: $d_{xy}=0.35J_1$, $J_z = J_2$, and NNN DMI as mentioned in the text.}
    \label{fig:patterns-Rh-bilayers-SKlattice}
    \end{figure}

In the spin patterns shown above for monolayers, Figure~\ref{fig:hyst-loops-15K}, and for bilayers,   Fig.\ref{fig:FM-DMIxy-XXX-Rh-15K-NNN}(b) and Fig.\ref{fig:hyst-loops-15K-bi-Rh-AFM}(b), only single skyrmion states have been found in the corresponding simulation processes for the assumed parameters. Of course, for other parameters, especially for stronger DMI, one can achieve stable multi-skyrmion textures or skyrmion lattices~\cite{Behera_skyrmmion,xu_magnetic_2020}. However, we show now that one can also create arrays of (quasi)stable skyrmions by appropriate artificial arrangement of initial spin states, by reversing manually the local magnetization at the nodes of a certain lattice, see the top panel in Fig.\ref{fig:patterns-Rh-bilayers-SKlattice}.  The parts (a) and (e) of this panel present random initial states, while the others are arranged manually in the magnetic field 0.3T (b-d) and  0.748T (f-h). The evolution of the initial states is analyzed by Monte Carlo simulations, and the second and third rows of these columns show the spin structure after  300k and 1M of Monte Carlo steps. 
In case (a) we get at the and (after 1M steps) individual as well as coupled skyrmions. In the case (b) the initially created square lattice evolves into a square lattice of skyrmions with a size smaller than the size of the initially created areas of reversed magnetization. In turn, in case (c) and (d) the system after 1M of simulation steps evolves to smaller skyrmions arranged in a defected hexagonal lattice.
The origin of the defects might be twofold. First, they appear naturaly, because of the square shape of the simulation cell, which is incompatible with a hexagonal lattice. Second, defects might be excited at increasing temperature.
The columns (e) to (h) in Fig.\ref{fig:patterns-Rh-bilayers-SKlattice} correspond to the same initial states as in columns (a) to (d), but simulations are in the magnetic field of 0.748T. In all the four cases (e) to (h), after 1M of Monte Carlo steps we get the arrays of skyrmions (of approximately equal size). Interestingly, the square lattice in column (f) survives after 1M of simulation steps. 
From the above results one may conclude that the simulated skyrmion lattices may be stable or quasi-stable in time, or can evolve into a disordered hexagonal lattice. This also shows that one can intentionally create skyrmion lattices, which may be of some interest in skyrmionics. However, their stabilization requires further investigations~\cite{stabilization}  

\subsection{Dynamical properties} 

\begin{figure}[h]
        \centering
        \includegraphics[width=\textwidth]{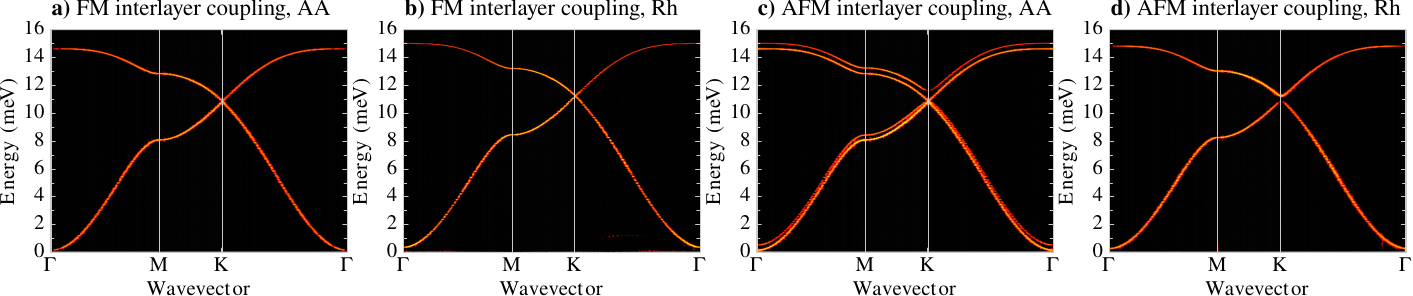}
        \caption{Dispersion curves of spin waves in bilayers with NNN DMI included (the corresponding parameters are given in the main text), and with NN DMI $d_{xy} = 0.05J_1$. Spin wave spectrum for the bilayer in the (a) AA stacking with FM interlayer coupling $J_z = J_2$; (b) Rh stacking with FM interlayer coupling $J_z = J_2$; (c) AA stacking with AFM interlayer coupling $J_z = -J_2$; (d) Rh stacking with AFM interlayer coupling $J_z = -J_2$. Other exchange parameters as given in the main text.} 
        \label{fig:Disp-AB}
    \end{figure}

Now, we consider briefly the spin excitations (spin waves) in both monolayers and bilayers of CrI$_3$. Therefore, we assume zero magnetic filed to exclude skyrmin texture formation when DMI is relatively large. First, we consider spin waves in bilayers assuming intrinsic NNN DMI parameters as given in the main text. As for the NN DMI we assume rather small its component $d_{xy}$, i.e., $d_{xy} = 0.05J_1$, which may be induced externally, e.g, by a gate voltage or strain due to a substrate, while for the second component we assume $d_z=0$. We focus on spin waves in bilayers in the AA and Rh stackings and with FM and AFM interlayer exchange coupling. 
The corresponding dispersion curves for the  bilayers with FM and AFM interlayer exchange coupling are shown in   Fig.\ref{fig:Disp-AB}.  The dispersion curves have been calculated using ASD Vampire code~\cite{Vampire_MonteCarlo}, and  are shown there  along the main paths in the Brillouin zone. 
Within this approach, dynamics of individual spins is described in terms of Landau-Lifshitz-Gilber equation. To determine the spin wave mode, the system is externally excited and then Fourier analysis of the time evolution is used to determine the spectrum. 
As in a monolayer there are two modes~\cite{kartsev_2020,jin_raman_2018,aguilera_topological_2020,zhang_gate-tunable_2020,hidalgo-sacoto_magnon_2020}, one can expect four modes in bilayers due to splitting induced by the exchange coupling between the two individual monolayers. However, the interlayer coupling is rather small, so the splitting is not well resolved. 
Indeed, in the case of FM interlayer coupling, the splitting of the modes are not resolved in  Figs\ref{fig:Disp-AB}(a,b) for the AA and Rh stackings. The splitting is clearly resolved in the case of AFM interlayer coupling in the AA stacking, see Fig.\ref{fig:Disp-AB}(c), while it is not well resolved in the Rh stacking, see Fig.\ref{fig:Disp-AB}(d) However, in the latter case there is a small gap in the spectrum, induced by the NNN DMI. However, this gap is very small  due to small values of NNN DMI parameters. To check variation of this gap with NNN DMI  we consider the spin wave spectrum in the monolayer case, where the NNN DMI  can be significantly enhanced, e.g., in the corresponding chromium trihalide Janus structures~\cite{xu_magnetic_2020}.        

\begin{figure}[h]
    \centering
    \includegraphics[width=\textwidth]{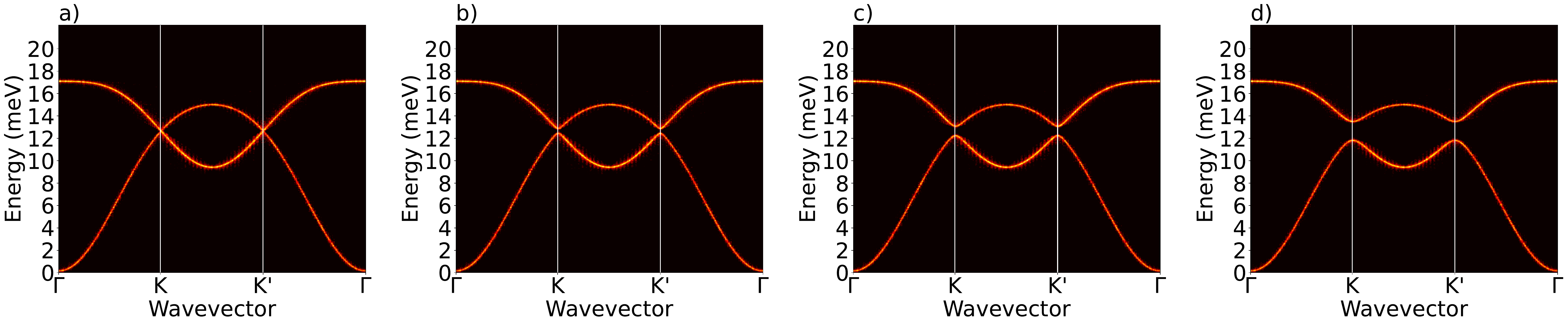}
    \caption{Dispersion curves of spin waves in a monolayer (a) without DMI, and (b) -- (d) with NNN DMI. The NNN DMI parameters used in the calculations: $d_{xy}^{\rm NNN} = -0.146\times \eta\, {\rm meV}$, 
    $d_z^{\rm NNN} = - 0.0176\times \eta\ {\rm meV}$, where
    (b) $\eta = 1$, (c) $\eta = 2$, and (d) $\eta = 4$. The other parameters:
    $J_1 = 1.224\, {\rm meV}$, $J_2 = 0.1986\, {\rm meV}$, and    $J_3 = 0.0027\, {\rm meV}$.}
    \label{fig:Disp-mono}
\end{figure}

Now we analyse spin waves in  monolayers~\cite{kartsev_2020,jin_raman_2018,aguilera_topological_2020,zhang_gate-tunable_2020,hidalgo-sacoto_magnon_2020}, assuming  relatively large values of the NNN DMI parameters, and focus on the topological features in the spectrum  associated  with NNN DMI. Though this term in DMI is small in pristine CrI$_3$, it can be remarkably enhanced, e.g., in the chromium trihalide Janus structures~\cite{xu_magnetic_2020}, to produce remarkable effects in the spin wave spectrum by creating a gap in the Dirac point K of the Brillouin zone. 
The  spin wave modes for monolayers are simulated within the ASD Uppsala code~\cite{Skubic_2008} . As for exchange parameters, we note, that these parameters are not well determined yet and there is some scatter in the available data~\cite{Pizzochero_2020,jaeschke-ubiergo_theory_2021,kartsev_2020}. We assume here the parameters consistent with those provided in Ref.~\cite{kartsev_2020}. 
The results of simulations are shown in Fig.\ref{fig:Disp-AB} for four different values of the NNN DMI parameters.
In the absence of NNN DMI, there is no gap in the spectrum. However, with increasing DMI parameters the gap opens in the spectrum at the Dirac point K of the Brillouin zone, and increases with increasing parameters of NNN DMI.  

\section{Summary and conclusions}

In this paper we have studied magnetic properties of chromium trihalides using numerical simulations within the atomistic spin dynamics and Monte Carlo  methods. These simulation techniques have been used to determine  critical  temperatures, hysteresis curves, and spin wave frequencies. These characteristics have been analysed for monlayers as well as bilayers in two (AA and rhombohedral) stackings and for ferromagnetic and antiferromagnetic interlayer exchange coupling. The main focus was on the influence of Dzyaloshinskii-Moriya interaction on the basic magnetic properties, especially on the hysteresis curves, spin textures (stripe domains, skyrmion textures, and others), and on the spin wave excitations, where Dzyaloshinskii-Moriya coupling between next-nearest-neighbours leads to opening a gap in the spectrum at the Dirac K point. Both, skyrmion formation and gap opening require quite a relatively strong DMI between nearest-neighbours and next-nearest-neighbours. Accordingly to observe these features one needs enhance the DMI either by external electric field (or strain), or use chromium trihalide Janus structures~\cite{xu_magnetic_2020}.

\bibliography{Ref.bib}



\section*{Acknowledgements (not compulsory)}

This work has been supported by the Norwegian Fi- nancial Mechanism 2014–2021 under the Polish Nor- wegian Research Project NCN GRIEG 2Dtronics no. 2019/34/H/ST3/00515.

\section*{Author contributions statement}

S.S. performed numerical simulations using Vampire code. P.B. performed numerical simulations using ASD Uppsala code. M.J. supported numerical calculations of S.S.. A.D. conceived of the idea. A.D and J.B supervised the work and wrote the manuscript with
input from all authors. 




\end{document}